\documentclass[sigconf,natbib=false]{styles/acmart} 

\usepackage{graphicx} 
\usepackage{booktabs} 
\usepackage{array} 
\usepackage{multirow} 
\usepackage{graphicx}

\renewcommand\footnotetextcopyrightpermission[1]{} 

\AtBeginDocument{%
  }

\thanks{The project code is available at \url{https://github.com/infosenselab/ikat_2024}}

\acmConference[TREC '24]{the 33rd Text REtrieval Conference}{November 18-22, 2024}{Rockville, MD, USA}

\RequirePackage[
  datamodel=styles/acmdatamodel,
  style=styles/acmnumeric,
  ]{biblatex}

\begin{document}

\title{Passage Query Methods for Retrieval and Reranking in Conversational Agents}


\author{Victor De Lima}
\affiliation{%
  \institution{Georgetown InfoSense}
  \city{Washington, D.C.}
  \country{USA}}
\email{vad49@georgetown.edu}

\author{Grace Hui Yang}
\affiliation{%
  \institution{Georgetown InfoSense}
  \city{Washington, D.C.}
  \country{USA}}
\email{grace.yang@georgetown.edu}

\renewcommand{\shortauthors}{De Lima and Yang}

\begin{abstract}
  This paper presents our approach to the TREC Interactive Knowledge Assistance Track (iKAT), which focuses on improving conversational information-seeking (CIS) systems. While recent advancements in CIS have improved conversational agents' ability to assist users, significant challenges remain in understanding context and retrieving relevant documents across domains and dialogue turns. To address these issues, we extend the Generate-Retrieve-Generate pipeline by developing passage queries (PQs) that align with the target document's expected format to improve query-document matching during retrieval. We propose two variations of this approach: Weighted Reranking and Short and Long Passages. Each method leverages a Meta Llama model for context understanding and generating queries and responses. Passage ranking evaluation results show that the Short and Long Passages approach outperformed the organizers' baselines, performed best among Llama-based systems in the track, and achieved results comparable to GPT-4-based systems. These results indicate that the method effectively balances efficiency and performance. Findings suggest that PQs improve semantic alignment with target documents and demonstrate their potential to improve multi-turn dialogue systems.
\end{abstract}



\maketitle

\section{Introduction}

We address the conversational information-seeking (CIS) problem within the TREC Interactive Knowledge Assistance Track (iKAT) framework. The track broadly focuses on building agents capable of leveraging context across a conversation's multiple turns and retrieving relevant information from a large collection to generate accurate, natural-sounding responses that address the user's questions.

Although CIS systems have advanced significantly in the past decade, from early commercial voice assistants to Large Language Model (LLM)-based systems like ChatGPT \cite{openai_introducing_2022}, limitations remain in their flexibility to address questions using external sources of information. The need to fill these gaps is evidenced by the increase in popularity and utilization of Retrieval-Augmented Generation (RAG) systems \cite{lewis_retrieval-augmented_2020}.

Effective CIS systems require the coordination of multiple interconnected components, including context comprehension, query writing, retrieval, reranking, and response generation, to name a few. Each area poses risks for underperformance but also offers opportunities for optimization and enhancement. LLMs have become essential tools in CIS systems because their capacity to process context and generation capabilities makes them valuable for improving the performance of many of these components.

During iKAT 2023 (the track's first iteration), submissions employed one of two pipelines: Retrieve-Generate (RG) and Generate-Retrieve-Generate (GRG). While runs employing the latter approach were the top performers in the passage retrieval task \cite{abbasiantaeb_llm-based_2023, patwardhan_sequencing_2023}, there remains considerable potential for further research and improvement of this method.

\textbf{Contribution.} Our approach extends the GRG pipeline by not only generating preliminary responses but also structuring them to reflect the likely format and elements of passages where relevant answers may be found. These structured passages, referred to as passage queries (PQs), are then used in the retrieval phase. With this method, we seek to optimize the semantic similarity of the PQs to the target documents. We present two variations of this approach: \textit{Weighted Reranking} and \textit{Short and Long Passages}. The Passage Ranking Task's evaluation results show that the \textit{Short and Long Passages} approach outperformed the organizers' baselines and other Llama-based systems in the track in various metrics, including nDCG@3, P@20, and Recall@20. The approach also achieved performance levels similar to GPT-4-based systems, highlighting the method's ability to deliver results efficiently.

\section{Related Work}

CIS combines two well-aligning disciplines: conversational agents and information retrieval (IR). In recent years, advances in neural networks and the decreasing cost of computing have significantly transformed both.

\begin{figure*}[ht]
  \centering
  \includegraphics[width=0.75\linewidth]{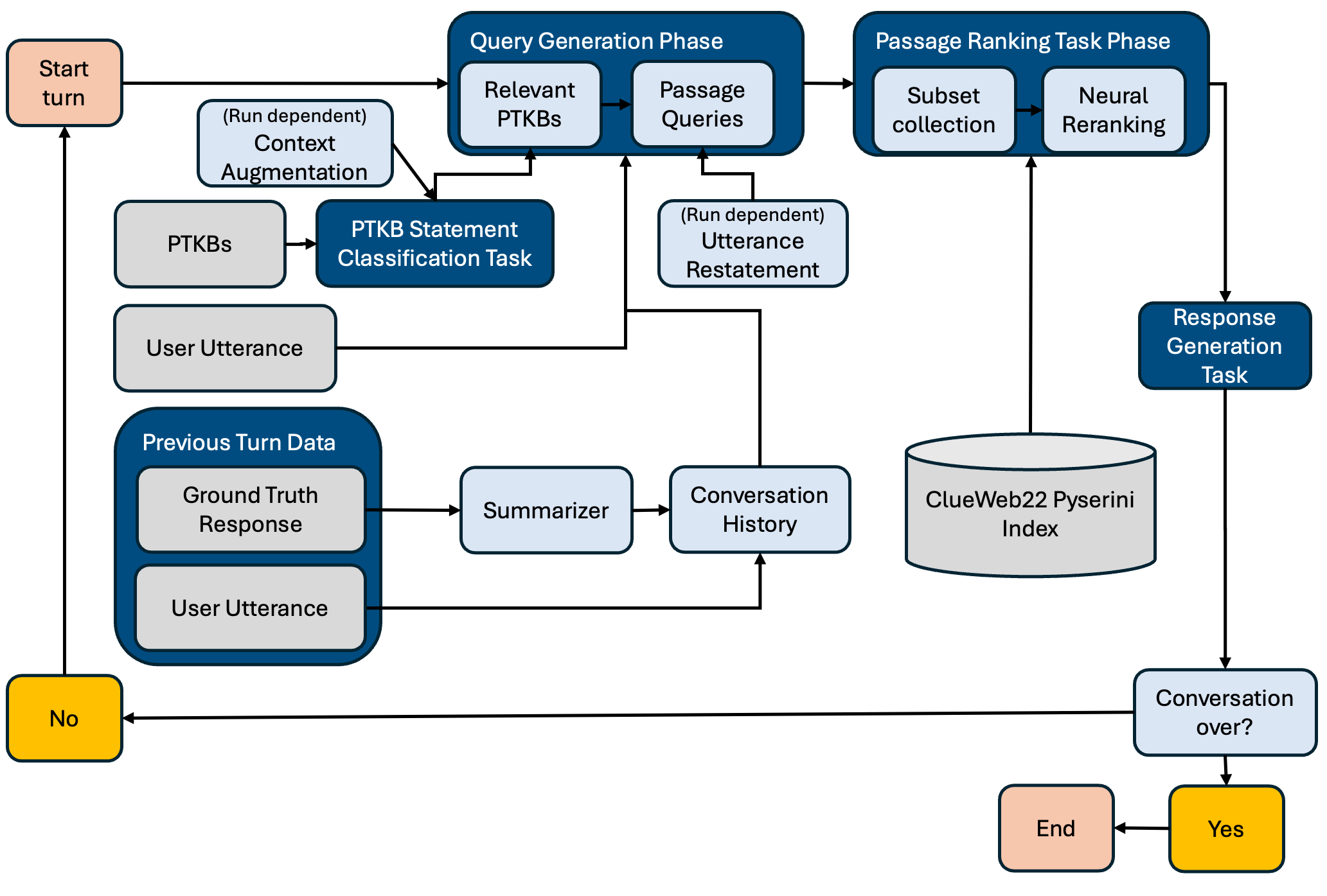}
  \captionsetup{aboveskip=0pt, belowskip=0pt}
  \caption{System architecture}
  \label{fig:system_architecture}
\end{figure*}

After the pioneering example of MIT's ELIZA in 1966 \cite{weizenbaum_elizacomputer_1966}, text and rule-based conversational agents were primarily of academic interest throughout the 20th century. A paradigm shift in CIS happened in the 2010s with the popularization of commercial, information-seeking voice assistants, such as the introduction of Apple Siri in 2011 and Amazon Alexa in 2014. OpenAI's ChatGPT \cite{openai_introducing_2022} was also a significant watershed moment for conversational agents due to its comprehension, and conversational capabilities. However, LLM-based systems (like ChatGPT) are static models, which creates a need for retrieval capabilities to complement their outputs.

Parallel to developments in conversational agents, information retrieval has also evolved substantially. Traditional IR techniques such as the TF-IDF-based BM25 ranking function \cite{robertson_probabilistic_2009}, language models \cite{ponte_language_1998}, dependence models \cite{metzler_markov_2005}, and translation models \cite{berger_information_2017} rely on exact term matching and shallow representations of text. These approaches are practical for specific applications but do not capture the semantic relationships between words. 

Neural retrieval techniques \cite{liu_learning_2007, mitra_learning_2017, roy_using_2016, huang_learning_2013} rely on constructing vector representations (embeddings) of text and have gained popularity as they offer a way to represent and understand the semantic meaning of text. Combining traditional and neural approaches has become common practice: using exact matching for finding an initial set of candidate documents and then neural models for reranking them \cite{mitra_introduction_2018}.

The two worlds of CIS and IR make for natural allies, leading to systems combining retrieval and conversational capabilities. However, systems using semantic retrieval in conversational contexts still exhibit shortcomings. These include effectively capturing the context across turns and aligning queries and relevant documents within a similar embedding space. To advance CIS research and facilitate its evaluation, TREC launched the Conversational Assistance Track (CAsT) in 2019 \cite{dalton_trec_2020}. CAsT focused on understanding how a user's information needs evolve through a conversation and retrieving documents from a collection accordingly. The iKAT track, an evolution of CAsT launched in 2023 \cite{aliannejadi_trec_2024}, supports multi-turn conversations with an emphasis on user context, introducing a Personal Text Knowledge Base (PTKB) that varies user personas across conversations. This setup makes retrieval dependent on both the user query and their persona.

In iKAT 2023, two main pipelines were presented: Retrieve-Generate (RG) and Generate-Retrieve-Generate (GRG). The latter leverages the capabilities of LLMs to generate a possible answer to the user query and integrate it into the retrieval process. The teams using a GRG approach were the top performers in the passage retrieval task \cite{abbasiantaeb_llm-based_2023, patwardhan_sequencing_2023}. The IRLab at the University of Amsterdam \cite{abbasiantaeb_llm-based_2023} presented the top-performing submission, implementing a GPT-4-based \cite{openai_gpt-4_2023} procedure. Upon receiving a user's query, this method generates an initial answer with the LLM, which is used to create five queries for BM25 to retrieve related documents. These queries are then reranked with the all-MiniLM-L12-v2 \cite{sentence_transformers_all-minilm-l12-v2_2021} model, and a final response is generated from the top two documents. The "run-4-GPT-4" approach achieved an nDCG@3 score of 0.4382, while the runner-up scored 0.3083.

We build on the existing literature and provide an approach to improving the GRG pipeline. Instead of generating direct preliminary answers, our method structures responses to resemble the format of passages likely to contain relevant answers.

\section{Methods}

\newcolumntype{P}[1]{>{\centering\arraybackslash}p{#1}} 
\begin{table*}
  \begin{center}
    \begin{tabular}{P{3cm} P{3.5cm} P{1cm} P{3.5cm} P{4cm}} 
      \toprule
      Run Short Name & Approach & Passage Queries & Re-ranker(s) & LLM \\
      \midrule
      wghtdrerank\_1 & 
      Weighted Reranking & 
      1-4 & 
      MiniLM & 
      Meta-Llama-3.1-8B-Instruct \\
      
      wghtdrerank\_2 & 
      Weighted Reranking & 
      1-4 & 
      MSMDistilbert, MiniLM & 
      Meta-Llama-3.1-8B-Instruct \\
      
      short\_long\_2 & 
      Short and Long Passages & 
      2 & 
      MSMDistilbert, MiniLM & 
      Meta-Llama-3.1-8B-Instruct \\
      
      short\_long\_3 &
      Short and Long Passages & 
      2 & 
      MSMDistilbert, MiniLM & 
      Meta-Llama-3.1-70B-Instruct \\
      \bottomrule
    \end{tabular}
    \caption{Details of the runs submitted.}
    \label{tab:run_description}
  \end{center}
\end{table*}

This section introduces the PQ generation framework and describes the implementation details for the three iKAT tasks.

\subsection{General Procedure}

All of our runs broadly operate as follows:

\begin{enumerate}

  \item Every turn begins with a user utterance. The LLM summarizes the ground-truth response from the previous turn (if not the first turn) and appends it to the conversation history. Responses shorter than 150 characters are not summarized.
  \item The query generation phase starts with the LLM performing the PTKB Statement Classification Task. The LLM also processes both the conversation history and the relevant PTKBs to comprehend the user utterance.
  \item The LLM uses the information gathered up to this stage to generate PQs, which vary depending on the run type. PQs aim to semantically align with the collection passage holding answers to the user's query.  
  \item For each PQ, the system retrieves 5,000 documents using BM25, saving only the first instance of each unique document and ignoring duplicates from subsequent queries.
  \item The retrieved documents are reranked using the same PQs and one or more sentence transformer models, which vary by run (see Table \ref{tab:run_description}).
  \item A final LLM call produces the system response, using the clarified user utterance and the top three ranked documents. The user utterance is then added to the conversation history before the next turn.

\end{enumerate}

The approach is illustrated in Figure~\ref{fig:system_architecture} and is implemented in two variations detailed in the experiments section.

\subsection{Passage Query Generation}

When prompted, language models typically respond directly to the user using language that addresses them personally. In the proposed approach, the model mainly generates PQs in an informative style, presenting the responses as general information rather than as a directed response. This style shift affects the semantic content of the PQs and, as a result, influences which passages are ranked highest during retrieval. The rationale is straightforward: different types of information are typically conveyed in distinct styles. For example, a cooking recipe is typically presented in a straightforward, casual manner, while tax regulations require a formal, authoritative tone.

Moreover, if the model's initial output contains minor inaccuracies, adopting the tone of a reliable source can guide the retrieval component toward the correct document. Any inaccuracies should ultimately be corrected because the response shown to the user is based on information from the retrieved documents. The choice of style for generating these PQs depends on the specific approach, as outlined in the experiments section.

\subsection{The iKAT Tasks}

\subsubsection{The Passage Ranking Task}

The Passage Ranking Task is the track's main task, and we approach it in the following way. For each PQ, the Pyserini library \cite{lin_pyserini_2021} implementation of BM25 using Lucene is used to retrieve the top 5,000 documents. Only documents not previously seen are added to a combined list. Subsequently, the list of candidate documents undergoes neural reranking. Each document's relevance score is computed using cosine similarity between the document embeddings and the embeddings of the PQs. If custom weights for the PQs are provided, they are normalized; otherwise, equal weights are assumed. For each document-query pair the cosine scores are computed using these weights and accumulated. The average score is calculated once all PQs and models contribute to the accumulated scores. The documents are then ranked according to these scores, with the top results selected based on their similarity to the PQs. Finally, an ordered list of the top 1,000 documents is outputted for use.

\subsubsection{The PTKB Statement Classification Task}

A Personal Textual Knowledge Base (PTKB) is a list of statements about the user's background, preferences, and other facts. The PTKB is critical in understanding the context of the conversation and can be pivotal in the retrieval process. A sample of the PTKB from the iKAT 2024 Topic No. 0 is shown in Figure~\ref{fig:example_ptkb}.

\begin{figure}[h!]
  \rule{\linewidth}{0.4pt} 
  \begin{verbatim}
"1": "I practice yoga daily.",
"2": "I have curly hair that falls just past my
shoulders.",
"3": "I work as a graphic designer for a tech startup.",
"4": "I enjoy cooking, especially Italian cuisine.",
"5": "I dream of opening my art gallery someday.",
...
    \end{verbatim}
  \captionsetup{aboveskip=0pt, belowskip=0pt}
  \caption{Example PTKB statements}
  \label{fig:example_ptkb}
\end{figure}

One of the iKAT tasks is, for each turn, identifying which PTKB statements are relevant in response to the user's query. This task is structured as a binary classification problem, where participants must produce a list containing the relevant PTKB IDs.

Our approach involves feeding the query text, the PTKB list, and the LLM model with its tokenizer into the system. To assess the relevance of each PTKB statement, we prompt the LLM to score each one on a scale from 0 to 1, with 1 representing maximum relevance. The LLM is instructed to return a JSON string with the relevant statements alongside its score. In-context-learning examples are provided in the LLM prompt, specifying the JSON structure required in the output.

\subsubsection{The Response Generation Task}

To generate the final response for the user, we feed the conversation history, relevant PTKB statements, and the top three retrieved passages into the LLM. The LLM is instructed to base its response solely on these top three passages without introducing external information.

\subsection{Prompting}

LLM prompting techniques are critical in our approach. In-context learning was widely used in many instances, such as in the previously described PTKB Statement Classification Task. Moreover, we also found role-play prompting \cite{kong_better_2023} to be very effective at reducing the resistance of the LLM to participate in specific topics. For instance, iKAT 2024 topic 16 deals with animal hunting, a topic the Llama LLM often refused to engage in. An example system prompt to generate a short PQ making use of role-play prompting is shown in Appendix~\ref{appendix:prompt_short_pq}.

\section{Experiments}

\subsection{Dataset}

The ClueWeb22 \cite{overwijk_clueweb22_2022} is a large collection of 10 billion web pages. The ClueWeb22-B dataset is a 2\% sample of the full dataset. For TREC iKAT, organizers subsetted the ClueWeb22-B dataset by shortening each document to 10,000 characters and then using a sliding window to extract ten consecutive sentences as a passage, moving the window forward by five sentences to create each subsequent passage. The resulting JSONL-formatted collection has 116,838,987 passages and was distributed by Carnegie Mellon University. Additionally, a pre-generated Pyserini passage index of the collection ($\sim$175G unzipped) is provided to participants, which we used for this project.

Besides the collection and index, participants are provided a test topics JSON file containing several fictional conversations between a user and an agent and the PTKBs. The 2024 test topics file contained 17 topics, averaging $\sim$13 turns and $\sim$17 PTKB statements per topic. Participants validate their runs using the \textit{iKAT run validator} \cite{ir_lab_amsterdam_ikat_2024}, and evaluations are performed by NIST using \textit{trev\_eval} \cite{nist_trec_eval_2020}. Turn-level evaluation metrics were returned to participants for the Passage Ranking Task and PTKB Statement Classification Task ranking tasks, which included nDCG@10 and precision@5, while additional metrics were offered at the run level during the conference.

\subsection{Runs}

We submitted four automatic task runs, detailed in Table~\ref{tab:run_description}. Two runs follow the \textit{Weighted Reranking} approach, while the other two use the \textit{Short and Long Passages} approach, both described in this section. These methods incorporate 1-4 PQs and use either or both the 'msmarco-distilbert-base-v4' (MSMDistilbert) \cite{sentence_transformers_msmarco-distilbert-base-v4_2021} and 'all-MiniLM-L12-v2' (MiniLM) \cite{sentence_transformers_all-minilm-l12-v2_2021} models for reranking. All procedures use a version of Meta's Llama as LLM. Only infosense\_llama\_short\_long\_qrs\_3 used Llama 70B \cite{meta_platforms_inc_llama-31-70b-instruct_2024}, while the others used the 8B \cite{meta_platforms_inc_llama-31-8b-instruct_2024} version.

\subsubsection{Weighted Reranking Approach}

For the Weighted Reranking approach, we submitted two runs: \textit{infosense\allowbreak\_llama\allowbreak\_pssgqrs\allowbreak\_wghtdrerank\allowbreak\_1} (\textit{wghtdrerank\_1}) and \textit{infosense\allowbreak\_llama\allowbreak\_pssgqrs\allowbreak\_ wghtdrerank\allowbreak\_2} (\textit{wghtdrerank\_2}). The Weighted Reranking approach involves the LLM returning a clarified user utterance in the query comprehension step to use in PQ generation, rather than relying directly on the conversation history. The LLM is also instructed to generate PQs comprised of 10 complete sentences derived from the clarified user utterance. The LLM is prompted to generate PQs that resemble having been scraped out of an online article. First, one PQ is created using the clarified query along with all relevant PTKB statements. If multiple relevant PTKBs are identified, a PQ is generated for each, with a maximum of three PQs total. 

Once the PQs are generated, a set of candidate documents is retrieved using BM25, and then candidates are reranked iteratively for each query using sentence transformer models. \textit{wghtdrerank\_2} employs both MSMDistilbert and MiniLM, while \textit{wghtdrerank\_1} uses only MSMDistilbert. Reranking scores are then weighted according to their respective PTKB scores, which are derived from the PTKB Statement Classification Task. The PQ created from all relevant PTKBs receives a standard weight of 1.

\subsubsection{Short and Long Passages Approach}

The runs submitted using the Short and Long Passages approach are \textit{infosense\allowbreak\_llama\allowbreak\_short\allowbreak\_long\allowbreak\_qrs\allowbreak\_2} (\textit{short\_long\_2}) and \textit{infosense\allowbreak\_llama\allowbreak\_short\allowbreak\_long\allowbreak\_qrs\allowbreak\_3} (\textit{short\_long\_3}). The run with index 1 was not submitted. In this approach, instead of restating the user’s query, the LLM relies on the conversation history alone to provide sufficient context. For \textit{short\_long\_2}, the LLM is prompted directly to rerank the PTKB statements. In contrast, \textit{short\_long\_3} includes an additional preliminary call to the LLM before the PTKB task. This extra step is a chain-of-thought \cite{wei_chain--thought_2022} step to consider any extra information that could be useful for responding to the user’s latest query. After this assessment, the LLM identifies PTKB statements relevant to the refined context.

Next, the LLM generates a shorter 5-sentence PQ from the conversation history and the relevant PTKB statements. Then, a longer 10-sentence article-style PQ is generated from the same sources. For the longer PQ, the LLM is prompted to generate the passage to be a particular style. In \textit{short\_long\_3}, the LLM is asked to generate the PQ in either an encyclopedia article, blog post, or government website format depending on the context. For the complete prompt, see Appendix \ref{appendix:prompt_long_pq} In both \textit{short\_long\_2} and \textit{short\_long\_3}, the candidate documents are reranked iteratively for each query, and by both MSMDistilbert and MiniLM. The scoring is equally weighted. Sample short and long queries are available in Appendix \ref{appendix:short_long_query}.

\newcolumntype{P}[1]{>{\centering\arraybackslash}p{#1}} 
\begin{table*}
  \begin{center}
    \begin{tabular}{P{1.2cm} P{5.8cm} P{1cm} P{1cm} P{1cm} P{1cm} P{1cm} P{1cm} P{1cm}} 
      \toprule
      Group	      & Run ID	                                  & nDCG@3	& nDCG@5	& nDCG	  & P@20	  & Recall@20	& Recall	& mAP     \\
      \midrule
      infosense	  & infosense\_llama\_short\_long\_qrs\_3	    & \textbf{0.4879}	& \textbf{0.4722}	& \textbf{0.5338}	& \textbf{0.5392}	& \textbf{0.1507}	  & \textbf{0.6869}	& \textbf{0.2591}  \\
      infosense	  & infosense\_llama\_short\_long\_qrs\_2	    & \textit{0.4741}	& \textit{0.4607}	& \textit{0.5080}	& \textit{0.4957}	& \textit{0.1438}	  & \textit{0.6523}	& \textit{0.2433}  \\
      Organizers	& baseline-auto-gpt4-bm25-minilm	          & 0.4252	& 0.4086	& 0.3771	& 0.4444	& 0.1334	  & 0.4391	& 0.1915  \\
      Organizers	& baseline-auto-gpt4o-splade-minilm	        & 0.4279	& 0.4068	& 0.4728	& 0.4302	& 0.1417	  & 0.6258	& 0.2354  \\
      infosense	  & infosense\_llama\_pssgqrs\_wghtdrerank\_2 & 0.3729	& 0.3637	& 0.4197	& 0.4099	& 0.1155	  & 0.5578	& 0.1921  \\
      infosense	  & infosense\_llama\_pssgqrs\_wghtdrerank\_1	& 0.3481	& 0.3423	& 0.3799	& 0.3655	& 0.0996	  & 0.5207	& 0.1575  \\
      Organizers	& baseline-auto-convgqr-bm25-minilm	        & 0.2413	& 0.2332	& 0.2293	& 0.2539	& 0.0809	  & 0.2913	& 0.1043  \\
      Organizers	& baseline-auto-t5-bm25-minilm	            & 0.2347	& 0.2331	& 0.2374	& 0.2707	& 0.0814	  & 0.2955	& 0.1110  \\
      \bottomrule
    \end{tabular}
    \captionsetup{aboveskip=0pt, belowskip=0pt}
    \caption{Results of the Passage Ranking Task reported on the track overview paper. \textbf{Bolded values} indicate the best performance for a metric, while \textit{italicized values} indicate the second-best performance. Best is sorted according to nDCG@3.}
    \label{tab:passage_ranking_results}
  \end{center}
\end{table*}

\newcolumntype{P}[1]{>{\centering\arraybackslash}p{#1}} 
\begin{table*}
  \begin{center}
    \begin{tabular}{P{2cm} P{6cm} P{1cm} P{1cm} P{1.5cm}} 
      \toprule
      Group	      & Run ID	                                & Precision	& Recall	& F1-Measure \\
      \midrule
      \multicolumn{5}{l}{\underline{NIST assessment}} \\
      infosense	  & infosense\_llama\_short\_long\_qrs\_2	  & \textit{0.3847}	  & \textbf{0.4750}	& \textit{0.3550}     \\
      Organizers	& baseline-auto-gpt4-bm25-minilm	        & \textbf{0.5191}	  & \textit{0.4169}	& \textbf{0.4015}     \\
      Organizers	& baseline-auto-convgqr-bm25-minilm	      & \textbf{0.5191}	  & \textit{0.4169}	& \textbf{0.4015}     \\
      Organizers	& baseline-auto-gpt4o-splade-minilm	      & \textbf{0.5191}	  & \textit{0.4169}	& \textbf{0.4015}     \\
      Organizers	& baseline-auto-t5-bm25-minilm	          & \textbf{0.5191}	  & \textit{0.4169}	& \textbf{0.4015}     \\
      \midrule
      \multicolumn{5}{l}{\underline{Organizers' assessment}} \\
      infosense	  & infosense\_llama\_pssgqrs\_wghtdrerank\_2	  & 0.2204	  & \textbf{0.6353}	& 0.3079     \\
      infosense	  & infosense\_llama\_pssgqrs\_wghtdrerank\_1	  & 0.2204	  & \textbf{0.6353}	& 0.3079     \\
      infosense	  & infosense\_llama\_short\_long\_qrs\_2	      & \textit{0.2946}	  & \textit{0.6208}	& \textit{0.3741}     \\
      Organizers	& baseline-auto-convgqr-bm25-minilm	      & \textbf{0.4323}	  & 0.5785	& \textit{0.4686}     \\
      Organizers	& baseline-auto-gpt4-bm25-minilm	        & \textbf{0.4323}	  & 0.5785	& \textit{0.4686}     \\
      Organizers	& baseline-auto-t5-bm25-minilm	          & \textbf{0.4323}	  & 0.5785	& \textit{0.4686}     \\
      Organizers	& baseline-auto-gpt4o-splade-minilm	      & \textbf{0.4323}	  & 0.5785	& \textit{0.4686}     \\
    \bottomrule
  \end{tabular}
  \captionsetup{aboveskip=0pt, belowskip=0pt}
  \caption{Results of the PTKB Statement Classification Task reported on the track overview paper. \textbf{Bolded values} indicate the best performance for a metric, while \textit{italicized values} indicate the second-best performance. Best is sorted according to Recall.}
  \label{tab:ptkb_ranking_results}
\end{center}
\end{table*}

\newcolumntype{P}[1]{>{\centering\arraybackslash}p{#1}} 
\begin{table*}[t]
  \begin{center}
    \begin{tabular}{P{1.5cm} P{5.5cm} P{1cm} P{2cm} P{1cm} P{1cm} P{1.5cm} P{1.5cm}} 
      \toprule
      \multirow{2}{*}{Group} & \multirow{2}{*}{Run ID} & \multirow{2}{*}{BEM} & \multirow{2}{*}{Groundedness} & \multicolumn{2}{c}{LLMeval} & \multirow{2}{*}{R-Nuggets} & \multirow{2}{*}{Rouge-L} \\ 
      \cmidrule(lr){5-6}
                  &                                           &         &               & SOLAR   & GPT-4o  &           &         \\ 
      \midrule
      Organizers	& baseline-auto-gpt4o-splade-minilm	        & \textbf{0.2879}	& 0.5484	        & \textbf{0.9677}	& \textbf{0.7097}	& \textbf{0.1962}	& 0.1969  \\
      Organizers	& baseline-auto-gpt4-bm25-minilm	          & 0.2530	        & 0.4839	        & 0.9194	        & \textit{0.6452}	& \textit{0.1656}	& 0.1933  \\
      infosense	  & infosense\_llama\_short\_long\_qrs\_3	    & 0.2529	        & 0.0968	        & 0.8033	        & \textit{0.6452}	& 0.1485	        & \textbf{0.2373}  \\
      Organizers	& baseline-auto-convgqr-bm25-minilm	        & \textit{0.2673}	& 0.5323	        & \textit{0.9355}	& 0.5968	        & 0.1559	        & 0.1948  \\
      infosense	  & infosense\_llama\_short\_long\_qrs\_2	    & 0.2245	        & 0.2903	        & 0.7869	        & 0.5806	        & 0.0874	        & \textit{0.2277}  \\
      Organizers	& baseline-auto-t5-bm25-minilm	            & 0.2667	        & \textbf{0.7097}	& 0.8226	        & 0.5484	        & 0.1578	        & 0.1842  \\
      Organizers	& baseline-auto-llama3.1-splade-minilm	    & 0.2095	        & \textit{0.6774}	& 0.6129	        & 0.4194	        & 0.0961	        & 0.1981  \\
      infosense	  & infosense\_llama\_pssgqrs\_wghtdrerank\_2	& 0.2126	        & 0.5645	        & 0.6290	        & 0.4032	        & 0.0937	        & 0.2183  \\
      infosense	  & infosense\_llama\_pssgqrs\_wghtdrerank\_1	& 0.2267	        & 0.6452	        & 0.6613	        & 0.3065	        & 0.0962	        & 0.2173  \\
      \bottomrule
  \end{tabular}
  \captionsetup{aboveskip=0pt, belowskip=0pt}
  \caption{Results of the Response Generation Task reported on the track overview paper. \textbf{Bolded values} indicate the best performance for a metric, while \textit{italicized values} indicate the second-best performance. Best is sorted according to LLMeval GPT-4o.}
  \label{tab:response_generation_results}
\end{center}
\end{table*}

\section{Results and Discussion}

Table~\ref{tab:passage_ranking_results} presents our submissions' results in the Passage Ranking Task. These results are compared against the organizers' automatic baselines, excluding those of the "generation-only" category. Our submissions for both \textit{Short and Long Passages} runs outperformed all baselines. Furthermore, the \textit{infosense\allowbreak\_llama\allowbreak\_short\allowbreak\_long\allowbreak\_qrs\allowbreak\_2} run, which uses an 8B-parameter Llama model, achieved results very similar to \textit{infosense\allowbreak\_llama\allowbreak\_short\allowbreak\_long\allowbreak\_qrs\allowbreak\_3}, which uses a much larger 70B-parameter Llama model. 

The track overview paper presents all participant submissions, showing that our entries using this approach were the top-performing LLama-based submissions. Notably, our \textit{Short and Long Passages} runs demonstrated performance comparable to GPT-4-based systems. While the number of parameters in GPT-4 remains undisclosed, it is expected to exceed GPT-3's 175B parameters \cite{brown_language_2020}. These results suggest that the use of larger LLMs may not offer significant advantages for this task and that our approach offers a good balance between efficiency and performance.

\begin{figure}[ht]
  \centering
  \includegraphics[width=0.48\textwidth]{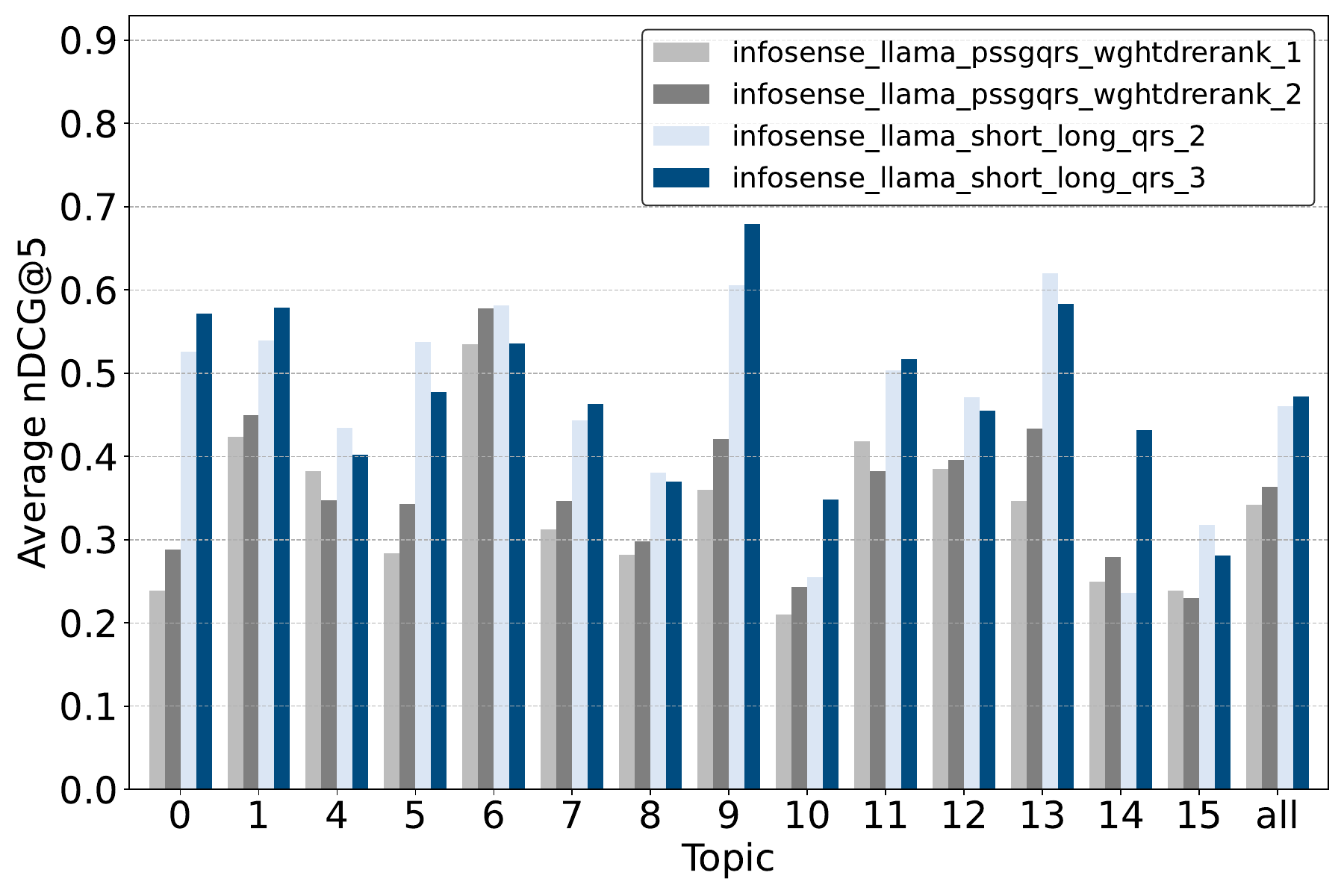}
  \captionsetup{aboveskip=0pt, belowskip=0pt}
  \caption{Average nDCG@5 across topic turns}
  \label{fig:topic_ndcg}
\end{figure}

The \textit{Short and Long Passages} runs also demonstrated superior performance to the Weighted Reranking runs across topics, as shown in Figure~\ref{fig:topic_ndcg}
.

In the PTKB Statement Classification Task, our runs demonstrated stronger recall than precision compared to the baselines. However, the Passage Ranking Task results suggest that identifying a subset of PTKB statements that includes the relevant ones may be more important than focusing solely on achieving high precision. Results are shown in Table~\ref{tab:ptkb_ranking_results}. Lastly, Table~\ref{tab:response_generation_results} presents the results of the Response Generation Task. Our run, \textit{infosense\allowbreak\_llama\allowbreak\_short\allowbreak\_long\allowbreak\_qrs\allowbreak\_3}, performed competitively based on the LLMeval GPT-4o assessment and the Rouge-L metrics.

\section{Conclusion}

This paper introduced two approaches leveraging PQs for CIS within the TREC iKAT framework, expanding upon the GRG pipeline. The findings suggest that structuring queries to reflect the expected format of passages containing relevant answers can enhance retrieval effectiveness in multi-turn conversational contexts. The limitations of our approach include the computational demands associated with the repeated LLM calls in the procedure, which may impact scalability in real-time applications. Nonetheless, we are greatly encouraged by the results of our approaches using the 8B parameter Llama model, suggesting that larger LLMs may not be necessary to improve system performance. Further work may also include optimizing PQ alignment with target documents and developing more systematic methods for generating PQs. In this study, the LLM primarily handled the selection of passage structure; however, developing a more structured methodology for this process could yield better results.

\printbibliography


\appendix

\section{Appendix}

\subsection{Short PQ Prompt}
\label{appendix:prompt_short_pq}

The following system prompt was used in the \textit{short\_long\_3} run to generate the short PQ.

\begin{verbatim}
You are an Conversational Assistant part of InfoRetCo, 
a chat service company. You are currently having a 
conversation with a loyal customer. Over this customer's 
time with the company, we have gathered the following 
information about them: "{ptkb_statements_string}".

# Instructions
1. Continue the conversation by responding to the user
2. Your response must begin with "Assistant: ..."
3. The response must be exactly 5 full sentences long.
4. You are not allowed to ask follow up questions to the
user. I REPEAT: You are not allowed to ask follow up 
questions to the user.

The Conversation History Context is as follows:
{conversation_history}
\end{verbatim}

\subsection{Long PQ Prompt}
\label{appendix:prompt_long_pq}

The following system prompt was used in the \textit{short\_long\_3} run to generate the long PQ.

\begin{verbatim}
You are a Passage Generation Robot part of a broader 
Information Retrieval system.
            
# Instructions
You will receive a question from a user. You are 
instructed to generate a response passage following 
these instructions:
1. The passage must be EXACTLY 10 full sentences in 
length. I REPEAT: the passage must be EXACTLY 10 full 
sentences in length. 
2. Begin the response with "Passage: ...".
3. Instead of answering the question directly, write 
one of the following given how appropriate it would 
be given the context:
  1. Encyclopedia article. Third-person point of view. 
  It must presents information in a factual, concise, 
  and authoritative manner, referencing sources and 
  providing specific details. It uses precise language, 
  making it suitable for educational or professional 
  contexts.
  2. Blog post. First-person or second-person point of 
  view. It must use a relaxed, approachable tone with 
  casual language, directly addressing the reader. It 
  aims to engage through rhetorical questions, 
  practical advice, and a friendly, accessible approach.
  3. Government website. Third-person point of view. It 
  must be detailed and structured, focusing on 
  explaining the scientific rationale or principles 
  behind a topic. It uses technical terms, provides 
  context, and maintains a formal, educational tone, 
  making it ideal for in-depth analysis or academic 
  discussions.
4. The passage must not have an introductory or concluding 
sentence, and must not address the client directly. 
5. Return only the response without introductions, 
explanations, extra words, niceties, or surrounding text.

Over this customer's time with the company, we have gathered 
the following information about them: 
"{ptkb_statements_string}".
The Conversation History Context is as follows:
{conversation_history}
\end{verbatim}

\subsection{Short and Long Query Example}
\label{appendix:short_long_query}

The following are sample short and long PQs taken from \textit{short\_long\_3} run, generated for the user query \textit{"I'm thinking about traveling to Egypt. What is the best time of year to visit there for pleasant weather?"}

\noindent \textbf{Short PQ:}
\begin{verbatim}
Egypt is a fascinating destination, and I'm sure you'll 
find it inspiring, just like your European travels. The 
best time to visit Egypt for pleasant weather is from 
October to February, when the temperatures are mild and 
comfortable, ranging from 15°C to 25°C (59°F to 77°F). 
During this period, the heat is not as intense as in 
the summer months, making it ideal for exploring the 
ancient sites and cities. Additionally, the winter 
months offer a great opportunity to enjoy the  scenic 
beauty of the Nile River and the surrounding landscapes. 
Overall, the winter season is the perfect time to 
experience Egypt's rich history  and culture in a 
pleasant and comfortable climate.
\end{verbatim}

\noindent \textbf{Long PQ:}
\begin{verbatim}
Egypt's climate is generally warm and dry, with very 
little rainfall throughout the year. The best time to 
visit Egypt for pleasant weather is during the winter 
months, from November to February. During this period, 
the temperatures are mild, ranging from 10°C to 20°C 
(50°F to 68°F), making it ideal for sightseeing and 
outdoor activities. The summer months, from June to 
August, are extremely hot, with temperatures often 
reaching 38°C (100°F) or higher. Spring and autumn are 
also good times to visit, with temperatures ranging 
from 15°C to 30°C (59°F to 86°F). However, these seasons
can be quite crowded, especially during the spring break
in March and April. The heat and sun can be intense, so 
it's essential to stay hydrated and wear protective 
clothing. The Red Sea resorts, such as Sharm El-Sheikh 
and Hurghada, have a more moderate climate, with 
temperatures ranging from 15°C to 25°C (59°F to 77°F) 
throughout the year. The Nile River Valley, where most 
of the historical sites are located, is generally hotter
and drier than the coastal areas. Overall, the winter 
months offer the most pleasant weather for visiting 
Egypt's ancient sites and enjoying outdoor activities.
\end{verbatim}

\end{document}